\documentclass[conference]{IEEEtran}

\IEEEoverridecommandlockouts
\usepackage{cite}
\usepackage{amsmath,amssymb,amsfonts}
\usepackage{algorithmic}
\usepackage{graphicx}
\usepackage{textcomp}
\usepackage{xcolor}
\usepackage{array}

\def\BibTeX{{\rm B\kern-.05em{\sc i\kern-.025em b}\kern-.08em
    T\kern-.1667em\lower.7ex\hbox{E}\kern-.125emX}}
\begin{document}
\begin{titlepage}
  \vspace*{1cm}
   J. Guan and A. Morris, "Extended-XRI Body Interfaces for Hyper-Connected Metaverse Environments," 2022 IEEE Games, Entertainment, Media Conference (GEM), St. Michael, Barbados, 2022, pp. 1-6, doi: 10.1109/GEM56474.2022.10017701.

       \vspace*{1cm}

       \copyright2022 IEEE. Personal use of this material is permitted.  Permission from IEEE must be obtained for all other uses, in any current or future media, including reprinting/republishing this material for advertising or promotional purposes, creating new collective works, for resale or redistribution to servers or lists, or reuse of any copyrighted component of this work in other works.

       \vspace{0.5cm}

       \vspace{1.5cm}
  
\end{titlepage}

\title{Extended-XRI Body Interfaces for Hyper-Connected Metaverse Environments

\thanks{Tri-council of Canada, Canada Research Chairs Program.}
}

\author{\IEEEauthorblockN{Jie Guan, Alexis Morris}
\IEEEauthorblockA{\textit{Adaptive Context Environments Lab} \\
\textit{OCAD University}\\
Toronto, Canada \\
\{jie.guan, amorris\}@ocadu.ca}
}

\maketitle

\begin{abstract}
Hybrid mixed-reality (XR) internet-of-things (IoT) research, here called XRI, aims at a strong integration between physical and virtual objects, environments, and agents wherein IoT-enabled edge devices are deployed for sensing, context understanding, networked communication and control of device actuators. Likewise, as augmented reality systems provide an immersive overlay on the environments, and virtual reality provides fully immersive environments, the merger of these domains leads to immersive smart spaces that are hyper-connected, adaptive and dynamic components that anchor the metaverse to real-world constructs. Enabling the human-in-the-loop to remain engaged and connected across these virtual-physical hybrid environments requires advances in user interaction that are multi-dimensional. This work investigates the potential to transition the user interface to the human body as an extended-reality avatar with hybrid extended-body interfaces that can interact both with the physical and virtual sides of the metaverse. It contributes: i) an overview of metaverses, XRI, and avatarization concepts, ii) a taxonomy landscape for extended XRI body interfaces, iii)  an architecture and potential interactions for XRI body designs, iv) a prototype XRI body implementation based on the architecture, v) a design-science evaluation, toward enabling future design research directions.
\end{abstract}


\section{Introduction}

    \begin{figure}[tbh]
 \centering 
 \includegraphics[width=\linewidth]{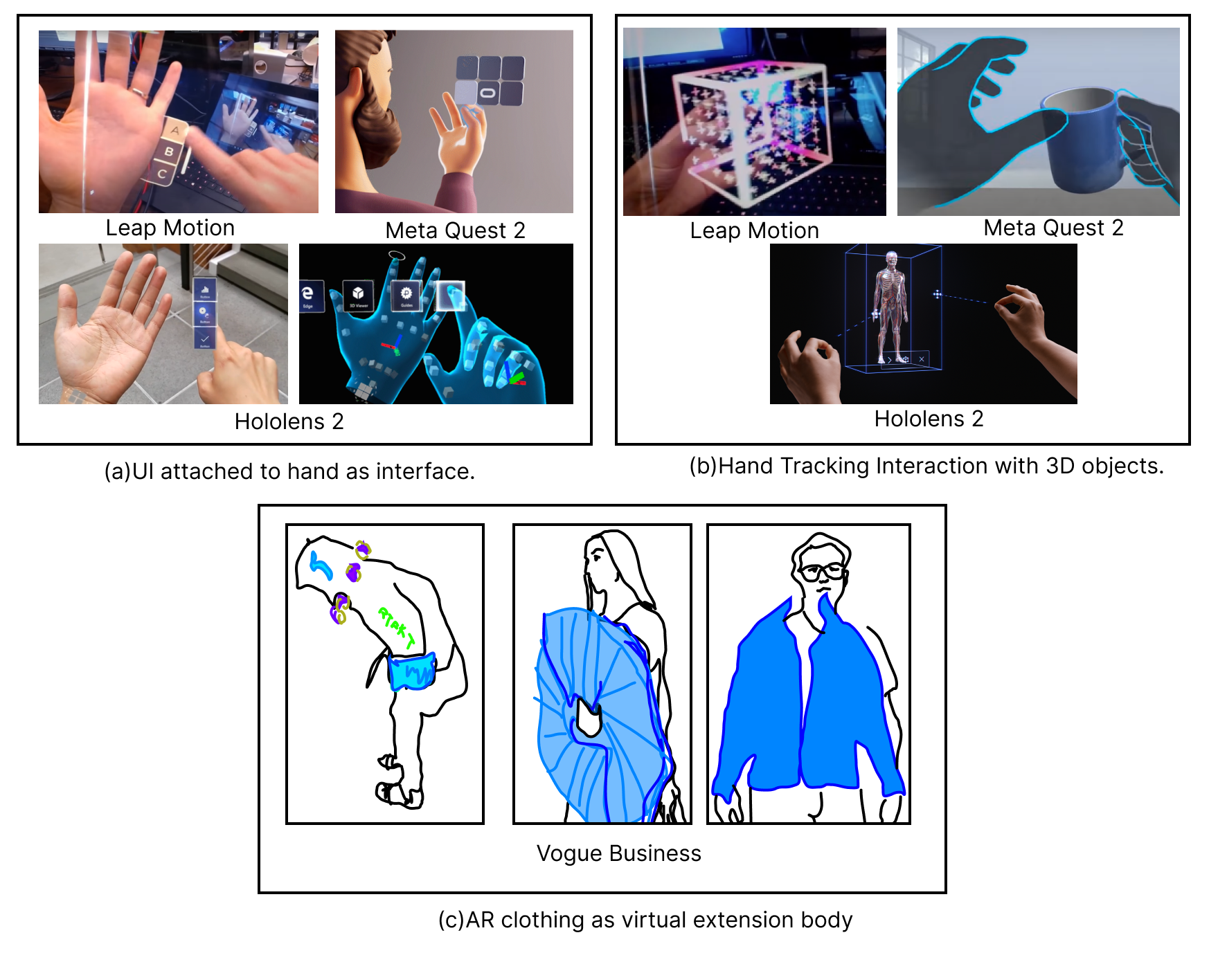}
 \caption{Metaverse interface designs include a) user interfaces attached to body with 2D UI elements \cite{northstarLeapMotion}\cite{microsoftHandMenu}\cite{oculusQeustHand}; b) 3D virtual objects with hand tracking interaction \cite{northstarLeapMotion}\cite{microsoftObjectInteraction}\cite{uploadvrOculusQuest}; and c) 3D mobile and wearable virtual clothing and avatars\cite{arclothvoguebusiness}. There is an opportunity to merge these toward an extended metaverse 3D body interface. 
}
 \label{handtracking}
\end{figure}

The metaverse refers to the set of technologies and ecosystems related to the merger of the virtual and physical environments and includes sensing elements as well as visualization for immersion. Thematically, the metaverse covers the shared dimensions of virtual worlds, augmented reality, mirror worlds or digital twins, and user and environmental sensing, as in \cite{cascio2007metaverse}\cite{ning2021survey}. In each case, the use of sensing components is essential, whether for tracking the user’s head orientation to their environments or for tracking other signals from the user, and the environment itself. These sensing elements may be centralized, such as within a user’s head-mounted display, or they may be embedded within the user’s environment in a decentralized fashion, or even across multiple environments. This data sensing elements bring the mixed reality domain closer to the realm of the internet-of-things, where edge device sensing, communication, and activation are essential capabilities. This is seen in multiple recent works that merge these technologies \cite{jo2019ar}\cite{jo2019iotAR}\cite{MorrisIsmarWorkstation}\cite{tsangtaraXRI}.

The metaverse, as a hybrid virtual-physical environment, requires a hybrid set of interactions that include both user interfaces and widgets, spatial references, as well as embedded visual effects, as in \cite{suzuki2022augmented}. As a result common 2D GUI design approaches may not be metaverse ready. For example, there has been recent approaches to provide metaverse interfaces attached to the body for hand interaction as well as AR clothing extensions. 

Figure \ref{handtracking}(a) shows design interfaces by Leap motion \cite{northstarLeapMotion}, Microsoft Hololens \cite{microsoftHandMenu}, Meta Quest 2 \cite{oculusQeustHand} based on these 2D GUI user interactions.  Likewise, Figure \ref{handtracking}(b) shows interactions with 3D objects via hand tracking (such as pinch and grab to pickup environmental objects)\cite{northstarLeapMotion}\cite{microsoftObjectInteraction}\cite{uploadvrOculusQuest} but are in-situ and not connected to the user’s body or wearable and mobile. Figure \ref{handtracking}(c) depicts embodied virtual clothing projects, such as \cite{arclothvoguebusiness}, which are wearable virtually and also mobile but do not provide interactive capabilities (whether virtual or physical). There is an opportunity to combine these distinct approaches toward a more engaging, embodied, interactive, and wearable metaverse  interaction through an extended XR-IoT body framework.

In this work, an extended body design is implemented with a virtual extension and wearable avatar interface for the user, as an early proof-of-concept for how to extend the human body in immersive metaverse environments. The contributions include: i) an exploration of metaverse, XRI, and avatarization concept, ii) a taxonomy landscape for extended XRI body interfaces, iii)  an architecture and potential interactions for XRI body designs, iv) a prototype of Extended-XRI body implementation based on the architecture, and v) a design-science evaluation of these artifacts.

This XRI extended body for metaverse research is presented as follows: Section II discusses the background theory of metaverse, XR-IoT and avatarization that relates to the Extended-XRI body. Section III presents a taxonomy landscape for Extended-XRI body. Section IV presents the architecture design, hybrid virtual and physical interaction in extended body, and a proof-of-concept prototype, and Section V presents a discussion and VI summarizes the paper.

\section{Background}
    The concept of an extended body for the metaverse and XRI requires exploration of multidimensional themes, specifically related to: i) metaverse theory, ii) XR-IoT theory,  and iii) avatarization theory. These themes are shown in table \ref{tableOfBackground}.


\begin{table*}[h!]
\centering
\begin{tabular}{|m{5em} |m{8em} |m{30em} |m{16em}|} 
 
 \hline
 Theory & Sources & Description & Relationship to Extended-XRI Body\\
 \hline
 Metaverse Theory 
 
 & Dionisio et al. (2013)\cite{dionisio20133d}, Lee et al.\cite{lee2021all}
 
 & Literally, "metaverse" is constructed by the prefix "meta"(beyond) and suffix "verse"(universe), that represents a world beyond the physical space \cite{dionisio20133d}. The definition of metaverse is diverse, and generally could be defined as the virtual environment with the state-of-art technologies such as internet, web technologies, and extended reality\cite{lee2021all}.

 & Metaverse as immersive virtual space provides the trajectory to consider the three-dimensional interface, and to consider virtual extension of body as interface.  \\

 \hline

 XRI Theory 
 
 & Jo \& Kim (2019)\cite{jo2019ar}, Morris et al. (2021)\cite{MorrisIsmarWorkstation}, Tsang \& Morris (2021)\cite{tsangtaraXRI}  
 & XR represents the virtual objects embody and overlay to the real environment, while IoT refers to the networking of objects \cite{jo2019ar}, the hybridization referred here as XR-IoT (XRI) \cite{tsangtaraXRI} to combine the IoT-based XR system and XR-based IoT system \cite{MorrisIsmarWorkstation}.

 & The XRI provides the perspective of connection of virtual and physical objects through IoT, and it could extend the connection of human hybrid virtual and physical body of the environments. \\

 \hline
 
 
 
 


 Avatarization Theory 
 &  Genay et al. (2021)\cite{genay2021being}
 & Avatarization continuum theory presents an augmented virtual body in various levels ranging from simple additional accessories to completely changing appearance and morphology. The continuum is from real body, body accesorization, partial avatarization, to the full avatarization. 
 
 & The avatarization provides the theory to consider the virtual extension embodiment of the body.  \\

 \hline

\end{tabular}
\caption{Core theories related to an extended body in the metaverse.}
\label{tableOfBackground}
\end{table*}


\subsection{Metaverse Theories}

Theories of the metaverse have been examined in \cite{dionisio20133d} and \cite{lee2021all} and summarized below. 
There are five stages (illustrated in \cite{guan2022extended}) of the metaverse evolution, identified by Dionisio et al.\cite{dionisio20133d}, starting from the text-based graphics world of the late 1970's to the open-source, decentralized, and co-creative platforms of today. The potential of the next stage of the metaverse may be toward a close connection with many real-life applications with immersive interface, to blur the border of the metaverse and physical world \cite{guan2022extended}. Today, the metaverse has various definitions since it is a state-of-the-art term that is being continually explored with emerging technologies. For example, Dionisio et al.\cite{dionisio20133d} presented that the metaverse is constructed by multiple individual virtual worlds, while Lee et al. \cite{lee2021all} considered the metaverse as a virtual environment with the state-of-art technologies such as internet, web technologies, and extended reality (XR).


\subsubsection{Extended Metaverse and Metaverse Disconnect\cite{guan2022extendingThesis}}

Based on a combination of theories including the MiRAs cube of \cite{holz2011mira}, and the reality-virtuality continuum of Milgram \cite{Milgram1994taxonomy}, the domain of an Extended Metaverse Agent is depicted extending along the dimensions of Mixed Reality Embodiment, Extended Interaction, and level of Agency. They represented how the virtual and physical entities can be presented on the hybrid space, and how they could interact with users. In this sense, the Extended Metaverse consists of one or more embodied virtual and physical objects, each having a degree of interactive properties in the virtual and physical dimension and having agent-oriented behavioral capabilities. The Extended Metaverse agent facilitates the cohesive connection of these agents with their real-world counterparts \cite{guan2022extendingThesis}.

\subsection{XR-IoT (XRI) Theory}

Augmented Reality (AR) and the Internet-of-Things (IoT) are key technologies receiving significant attention, as they improve the communication between the Metaverse and physical space. As examined in works like \cite{MorrisIsmarWorkstation} and \cite{tsangtaraXRI},  design and creation of a more connected mixed-reality environment necessitates the convergence of XR and IoT technology paradigms in order to connect with the physical environment while extending into the virtual environment. As discussed in \cite{jo2019ar}, Augmented Reality provides this through an interactive medium of overlaid virtual objects anchored to the real environment, whereas IoT refers to the networking of physical objects with computing devices for sensing and communication. 

This hybridization, as in \cite{tsangtaraXRI}, is referred to here as XR-IoT (XRI), and it represents the combination of XR-based IoT systems and IoT-based XR systems. \cite{MorrisIsmarWorkstation} \cite{morrisIoTAvatar} \cite{guanIoTAvatar2} \cite{shao2019iot} presented more comprehensive examples of immersive, information-rich, multi-user, and agent-driven XRI systems, developed by the authors. The combination of these technologies has the potential to bring humans and their environmental objects closer together, and future hybrid XRI applications for applied situations such as education, cyber security, and marketing\cite{andrade2019extended} are being developed. Much of these works are recent and gaining momentum as both paradigms attain maturity and adoption. Together these early explorations have set the stage toward the extended metaverse themes of this research.

\subsection{Avatarization and Embodiment Theories}

According to Avatar theory, \cite{bell2008toward}, \textit{``An avatar is any digital representation (graphical or textual), beyond a simple label or name, that has agency (an ability to perform actions) and is controlled by a human agent in real time.''} Avatars typically refer to the human body representation but can also be a virtual extended embodiment of physical objects in the physical environment, as in \cite{guanIoTAvatar2}\cite{morrisIoTAvatar}. 
Body Avatarization Continuum is a concept presented by \cite{genay2021being} and introduced augmented virtual body in various levels ranging from simple additional accessories to completely changing appearance and morphology. The continuum is from real body (represent the natural physical body) to the full avatarization (a completely virtual body). In between, there are body accessorization that describes the superficial elements (such as clothes or glass) attached to body and partial avatarization represents additional virtual body parts or removal of real ones as well as modification of the bodily characteristics. 
This paper centers on partial avatarization, as the virtual extension of the body. An example in this spectrum is using a virtual prosthesis to train a patient before implanting the real prosthesis \cite{soares2012virtual}. In addition, the virtual body in partial avatarization can not only replace or overlay the equivalent body, but also could modify the original structure and even extend the interaction possibility of the body. As an example, a virtually extended arm in augmented reality is presented by\cite{feuchtner2017extending} for users to manipulate the surroundings from a distance.

The above background sections highlight the overall need for more explorations that combine metaverse theory, XRI, and new forms of body avatarization and interaction.

\section{A Taxonomy Landscape for Extended XRI Body} 
 \begin{figure}[tbh]
 \centering 
 \includegraphics[width=\linewidth]{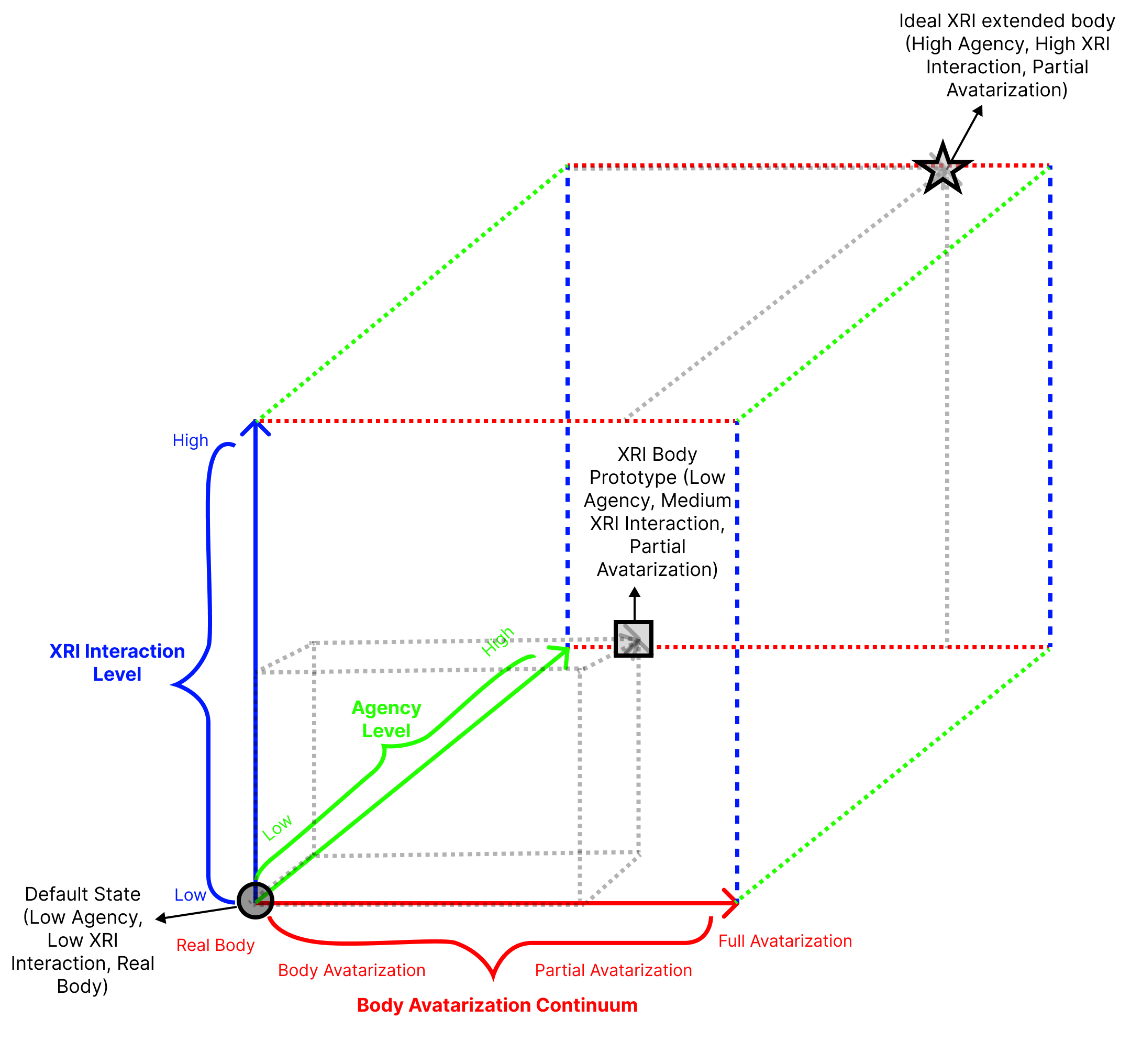}
 \caption{Extended-XRI body Taxonomy inspired by \cite{holz2011mira} \cite{genay2021being}. The circle represents the default state in the real world. The star represents the ideal XRI extended body interface.  The square represents the prototype in Section IV of this work.}
 \label{theory}
\end{figure}
The mixed reality agents (MiRAs) framework\cite{holz2011mira} presented the criteria of interactive capacity, corporeal presence, and agency level to define mixed reality agents that interact both physically and virtually. To extend the mixed reality agents concept, the human body is necessary for consideration as part of the mixed reality environment, with the virtual extension of the physical body as the three-dimensional virtual interface to interact and communicate with both the virtual and physical agents in context.

In order to define the XRI body field, it is, therefore, necessary to distinguish the physical and virtual embodiment (as the avatar) of the human body and the informational interaction in a mixed reality environment (see Figure \ref{theory}). For the body embodiment, Genay et al.\cite{genay2021being} presented a body avatarization continuum to extend the virtual representation in a virtual environment context. The real body is the material body that inhabits the physical environment, while the virtual body is the avatar to represent and be controlled by the real body. In between, there is the hybrid virtual and physical body with body avatarization and partial avatarization. 

The XRI interaction level represents how much information the Extended-XRI body could send and receive in order to interact with both the virtual and physical environments and the physical body itself, via an IoT communication broker. 

For agency, the theory from Wooldridge and Jennings \cite{wooldridge1995intelligent} provides the structure of what is low and high agency. In high agency, the agents should incorporate mentalistic notions, while low agency represents the basic capabilities of autonomy. Specific to the Extended-XRI body, it represents the level of how the virtual extension embodiment senses and reacts to both virtual and physical surroundings. In the metaverse context, as a combination of virtual-physical technologies, a metaverse agent must also be capable of performing hybrid virtual and physical agent interactions (i.e., autonomous, reactive, proactive, and social interactions \cite{wooldridge1995intelligent}).

\section{Extended-XRI Body for Hyper-Connected Metaverse Environments}


\subsection{Architecture Design}
 
Figure \ref{architectureInteraction} presents the hybrid-body-agent system as the user (wearing a mixed-reality device) in the human-in-the-loop hyper-connected smart metaverse environment, with virtual extension agents as body parts to communicate and interact with the virtual and physical surroundings (solid lines represent physical entities and dashed lines represent virtual entities). In the hybrid-body-agent system, the human body is augmented by virtual extension agents with the embodiment in mixed reality device and could sense and react to the context (to present agency) by the sensors attached to the physical body and on the environment. The hybrid environments contain both physical objects (with IoT-enabled devices and sensors) and virtual objects (the IoT Plant Avatar from\cite{morrisIoTAvatar} and other workstation objects from \cite{MorrisIsmarWorkstation}) that are visualized in Unity 3d with a mixed-reality device. The IoT communication broker component at the top contains three protocols (MQTT\footnote{https://mqtt.org/}, Socket.IO\footnote{https://socket.io/} and HTTP \footnote{https://developer.mozilla.org/en-US/docs/Web/HTTP/Overview} for adaptive different smart devices requirement) for sending messages between virtual and physical agents.

\subsection{Hybrid Virtual and Physical Interactions in Extended Body and Metaverse}
\begin{figure*}[tbh]
 \centering 
 \includegraphics[width=0.55\linewidth]{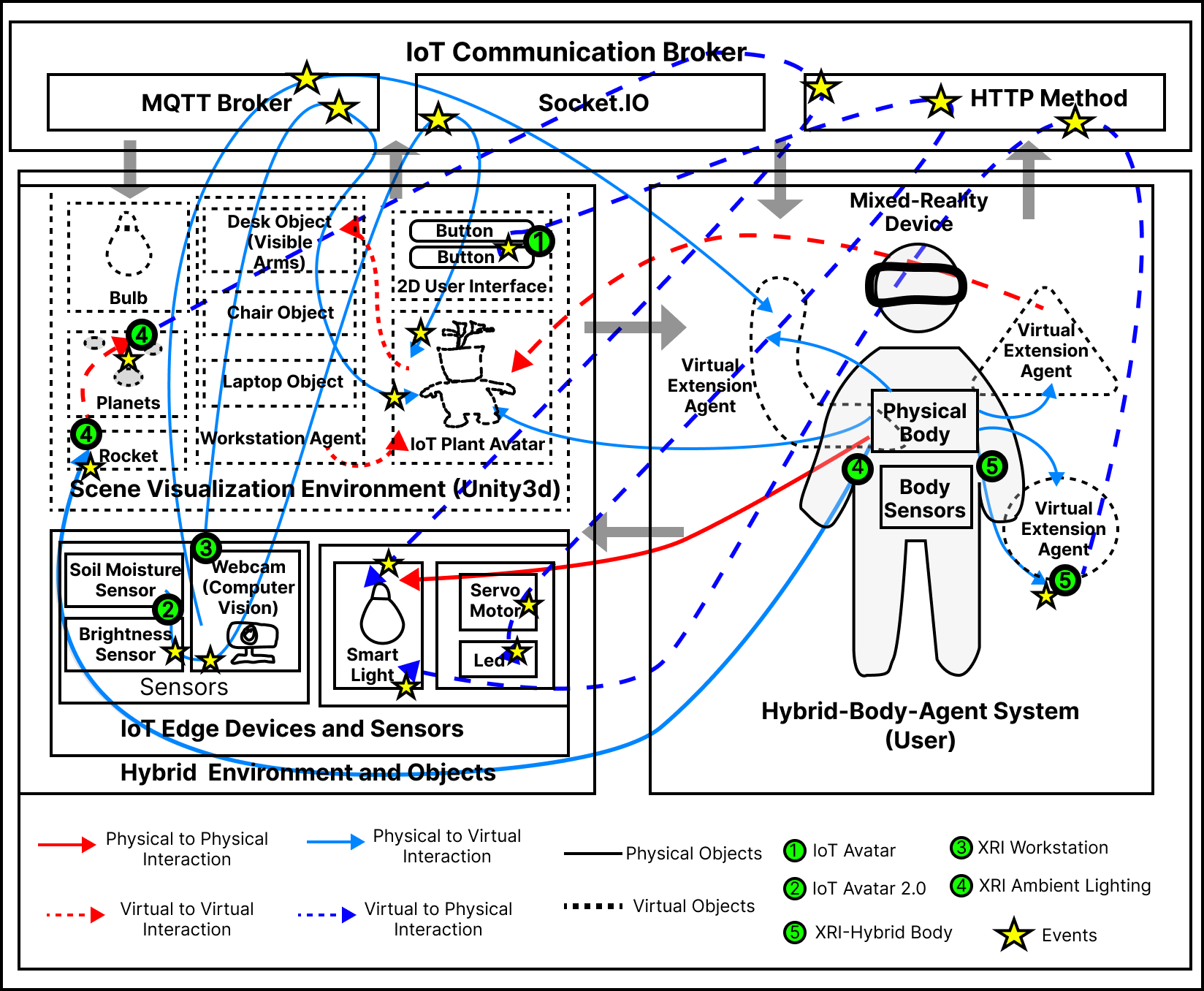}
 \caption{ A Human-in-the-Loop Hyper-Connected Smart Metaverse Environment Architecture with XRI-Hybrid Body. Users interact with the physical world containing IoT-enabled edge devices and sensors, as well as a scene visualization environment, and a communication broker. A mixed-reality device forms a hybrid body-agent system for the user, and four kinds of interaction are shown as directions for exploration; physical-to-physical, virtual-to-physical, physical-to-virtual, and virtual-to-virtual. Example projects from the authors' previous works are highlighted in the green circles, highlighting areas of exploration; IoT Avatar\cite{shao2019iot}, IoT Avatar 2.0\cite{guanIoTAvatar2} \cite{morrisIoTAvatar}, XRI Workstation\cite{MorrisIsmarWorkstation}, XRI Ambient Lighting \cite{guan2022extended}.}
 \label{architectureInteraction}
\end{figure*}

In terms of the interaction design (see Figure \ref{architectureInteraction}), this paper presents four way interaction with the hybrid-body-agent system in the mixed-reality environment. The solid red arrow is the \emph{default interaction} with physical body to interact with the physical object, for example, a user turns on or off a light with physical hand by pressing a physical button. 

The solid blue arrows represent the \emph{physical-to-virtual} connections, which can be through the user's direct interaction, captured by body tracking, or by other physical IoT-enabled devices that send messages via a communication broker. The user's physical body interaction is simulated 
using body-tracking hardware, such as Hololens 2 \cite{microsoftObjectInteraction}, Meta Quest 2 \cite{uploadvrOculusQuest} and Leap Motion \cite{northstarLeapMotion}, that track users' hands and provides them the ability to interact with virtual objects through gestures. In terms of the IoT interaction, it requires sensors setup in the environment or on the human body to send messages through the IoT Broker to the virtual objects. For example, the IoT Avatar 2.0\cite{morrisIoTAvatar} captures context information with a soil moisture sensor, brightness sensor, and webcam to control the emotions of a virtual plant avatar through the Socket.Io protocol. In addition, the XRI Workstation \cite{MorrisIsmarWorkstation} also used a webcam with a computer vision model to detect the on and off state of a lamp and whether or not a human is present in front of a laptop to control the scale and ambient effect of a virtual plant avatar.

The dashed red arrows indicate conventional \emph{virtual to virtual} interactions, referring to virtual objects and their behavior logic within a game engine (such as Unity). The dashed blue arrows show the interaction from \emph{virtual to physical}, which requires the IoT communication broker to perform information transfer. When an event or multiple events happen in the virtual environment, it could send message to the physical space to affect the physical objects. The initial IoT Avatar \cite{shao2019iot} is an early prototype with 2D buttons attached to the virtual plant avatar in Augmented Reality on a mobile phone. In this prototype, when a user pressed a button, it could trigger a servo motor and the LED light in the physical space using HTTP method as the communication protocol.

However, in some situations the interaction can not only be implemented with one way connection of virtual and physical, they require multiple interactions together. For example, the XRI Ambient Lighting in the extended metaverse framework \cite{guan2022extended} considered metaverse as the interface for the context. In this prototype, the user could pick up and move the rocket in the mixed reality (physical to virtual) to collide with the virtual planets (virtual to virtual) that could trigger an event to send message to the physical light to change the color of the smart bulbs through HTTP method (virtual to physical). This paper explored a XRI-hybrid body (also requires multiple ways interaction) that consider virtual extension body as the 3D interface in mixed reality. The physical body could interact (such as pinch, grab and poke) with the virtual extension (physical to virtual) to send a message through the IoT broker to the physical light to turn it on or off (from virtual to physical).

\subsection{An Early Extended XRI Body Prototype}

\begin{figure}[tbh]
 \centering 
 \includegraphics[width=\linewidth]{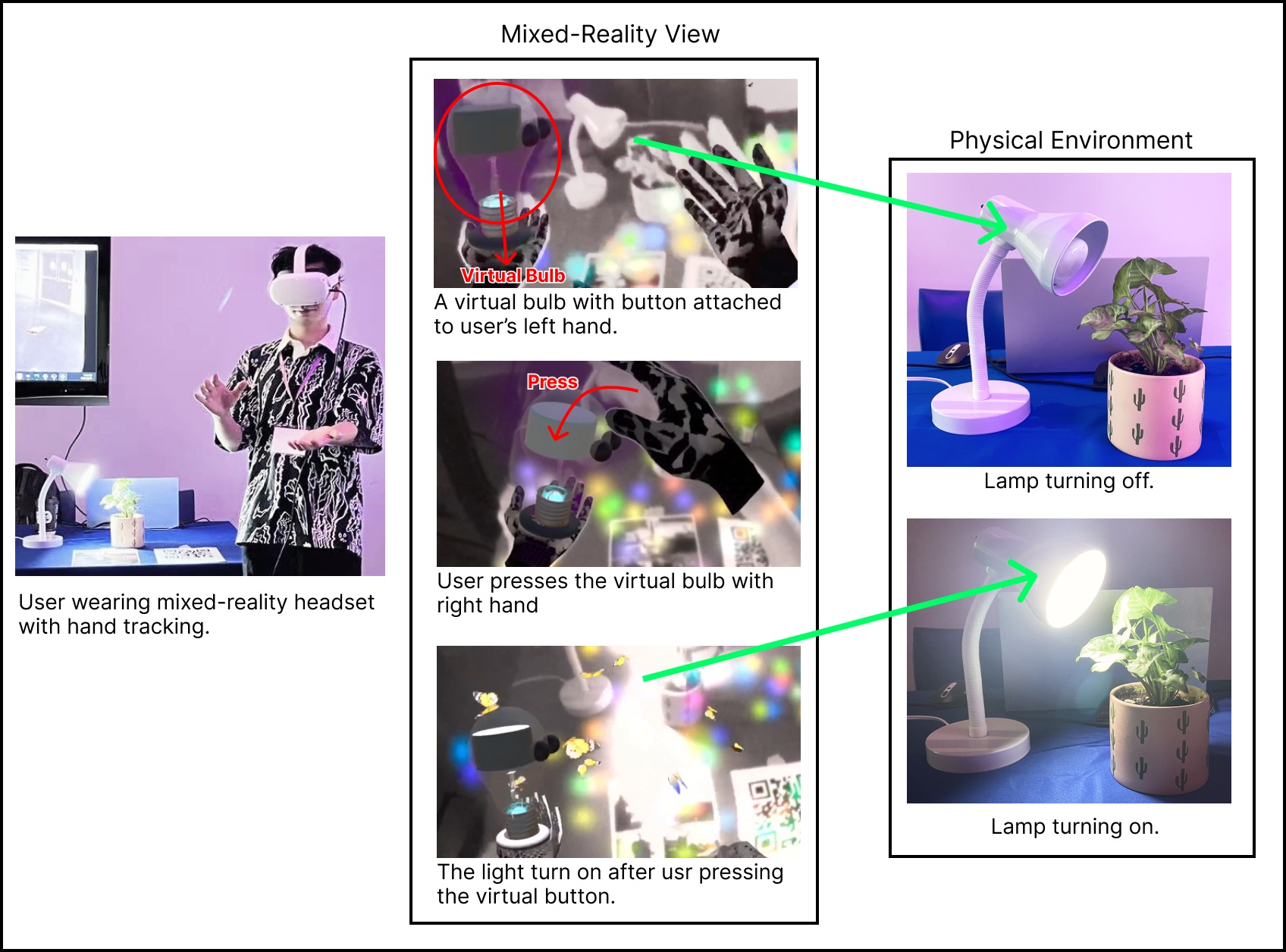}
 \caption{Prototype of extended human body in mixed-reality with virtual bulb attached to left hand. See interaction path 5 in Figure \ref{architectureInteraction}.}
 \label{prototype}
\end{figure}

Figure \ref{prototype} presents a prototype of the virtual extension body as interface to control the physical space.
As in Section III, the taxonomy landscape for Extended XRI Body, this prototype addressed low agency level, medium XRI Interaction degree, and located at Partial Avatarization. It is an early proof-of-concept demo to extend the human physical body virtually within the hyper-connected metaverse environment scenario. This approach requires the user to be wearing a Meta Quest 2 (as mixed-reality headset with pass-through enable) to see the hybrid virtual extension agent (virtual bulb) that is attached on the left hand with hand tracking. When the user presses the virtual bulb on their left hand, it will send messages from the unity3D build app in Meta Quest 2 that connects to Wifi to the physical lamp.  The physical lamp is the Philips Hue Bulb\footnote{https://www.philips-hue.com/en-sg/products/smart-lightbulbs} in the physical environment represents a IoT-enabled agent, to be controlled by the button press event with the body virtual extension by sending information with HTTP Put to turn on or off.

\section{Discussion}

This work fits into the design-science in information systems research framework \cite{hevner2004design}.
As such, it can be evaluated by following the design-science research guidelines. These are as follows:

In terms of 1) design as an artifact: 
This work has produced a viable taxonomy, architecture and implementation prototype for Extended-XRI Body. The taxonomy allows designers to understand where different XRI body designs fit within a single conceptual framework, versus the ideal design. The architecture is a synthesis of the authors' prior research prototypes into a holistic and general XRI system diagram that can inform future designs. The prototype presents a single XRI extended body interaction scenario (among others related to the architecture as a proof-of-concept).

In terms of 2) problem relevance: 
This work is relevant to both the metaverse disconnect needs for virtual-physical hyper-connectedness, as well as the human-factor needs (i.e., physical, psychological, social, organizational, political \cite{vicente2013human}), specifically the physical human-factor layer as it relates to the metaverse.

In terms of 3) design evaluation the focus is on the prototype artifact: 
As part of the design-science evaluation, a brief descriptive evaluation is presented for the Extended-XRI body for the case scenario below where a user controls a physical light bulb using a virtual extended body interface. In terms of the level of utility for the scenario this can be considered by the user being able to turn on and off the light regardless of their location or their context situation by using the extended body interface. The quality of the prototype can be considered based on factors like how reliably the user is able to control the light and the direct level of message passing to transmit the signal. The efficacy of the prototype can be considered based on ease of interaction when controlling the light and the embodiment of the bulb on the hand makes the interaction familiar to the user. 


In terms of 4) research contribution: 
Among the main contributions (conceptual, architectural, prototyping, evaluation) this work provides a new perspective on avatarization and mixed reality agents; a new synthesized architecture extracted from multiple XRI implementations as a reference design structure; a functional 3D interface attached to the body as an extension that is also an interface to the physical environment; and lastly, the use of design-science for metaverse prototype evaluation which indicates that the design process is valid for this research.



In terms of 5) research rigor: 
This paper used MiRAs\cite{holz2011mira}, Avatarization\cite{genay2021being}), and the standard development tools in the field (Unity and Philips Hue) for creating the prototype artifact and its evaluation \cite{hevner2004design}.

In terms of 6) design as a search process: 
Through the design of an Extended-XRI body to connect the metaverse and physical space and address the human factor needs for the metaverse, this work has produced a new overall method for XRI body systems, toward the  metaverse disconnect problem and metaverse human factors.

In terms of 7) communication of research: This work is still in the early stages, and it is being explored by the community in the current paper as well as the author's previous papers. In future work, this would be extended through more prototypes including i) Workstation Agents as body extensions for adaptive scenarios related to physical contexts; ii)XRI-hybrid body for multi-user scenarios; iii)XRI-hybrid body in large-scale extended metaverse environments, and using the presented Extended-XRI taxonomy to analyze these potential works and other related work.

\section{Summary}

The mixed reality and Internet of Things domains are converging alongside other metaverse technologies toward immersive and adaptive smart environments. This brings with it new forms of user interaction and new challenges that are multimodal and also multidimensional, as users must engage with both physical and virtual objects and interfaces to achieve their goals, tasks, etc. This work contends that more is needed to extend the users body into the immersive environment and that the user is central to the interaction within this environment. Extended body component widgets have been explored in this work and the key outcomes include: opportunities for new forms of virtual- physical body interface and a taxonomy for extended body concepts derived from the literature; a general architecture for XRI system interactions that include extended body systems; a prototype extended body interaction for the light control scenario; and a design science evaluation of this approach and its resulting artifacts. It is hoped that this work will help future researchers as they consider new forms of hybrid user interaction in immersive adaptive environments toward the coming metaverse.

\section*{Acknowledgment}
This work was supported by funding from the Tri-council of Canada under the Canada Research Chairs program.

\bibliographystyle{IEEEtran}
\bibliography{references}

\end{document}